\newif\ifaps\apstrue
\newif\ifnewrevtex\newrevtexfalse
\newif\ifiop\iopfalse

\ifaps
\ifnewrevtex\documentclass[english, aip, pof, amssymb, amsmath, preprint]{revtex4-2}
\else\documentclass[english, aip, pof, amssymb, amsmath, preprint]{revtex4-1}
\fi
\usepackage{dcolumn}
\usepackage{bm}
\fi

\ifiop
\documentclass[]{iopart}
\usepackage{iopams}
\expandafter\let\csname equation*\endcsname\relax
\expandafter\let\csname endequation*\endcsname\relax
\usepackage{amsmath}
\usepackage{amssymb}
\fi

\usepackage[english]{babel}
\usepackage{graphicx}
\usepackage{verbatim}
\usepackage{hyperref}
\usepackage{breakurl}

\begin{document}

\ifaps
\title{Approximate analytic solution of the potential flow around a rectangle}
\author{Eunice J. \surname{Kim}}
\affiliation{Microsoft Corp., Redmond, WA 98052}
\author{\surname{Kim} Ildoo}
\email{ildoo.kim.phys@gmail.com}
\affiliation{FG Research LLC, Bellevue, WA 98004}
\date{\today}
\fi

\begin{abstract}
In undergraduate classes, the potential flow that goes around a circular cylinder is designed for complemental understanding of mathematical technique to handle the Laplace equation with Neumann boundary conditions and the physical concept of the multipolar expansion.
The simplicity of the standard problem is suited for the introductory level, however, it has a drawback.
The discussion of higher order multipoles is often missed because the exact analytic solution contains only the dipole term.
In this article, we present a modified problem of the potential flow around a rectangle as an advanced problem.
Although the exact solution of this case is intractable, the approximate solution can be obtained by the discretization and the optimization using multiple linear regression.
The suggested problem is expected to deepen the students' insight on the concept of multipoles and also provides an opportunity to discuss the formalism of the regression analysis, which in many physics curricula is lacking even though it has a significant importance in experimental physics.
\end{abstract}

\ifaps
\maketitle
\fi

\section{Introduction}

One cannot stress enough the importance of the Laplace equation in physics.
Any irrotational and solenoidal vector field, that is, all conservative field in free space, is a gradient of a scalar function, and such defined potential function satisfies the Laplace equation. 
Due to its wide applicability in nature, the Laplace equation is frequently discussed in undergraduate curricula.
Examples of the problem sets include the electric field inside a rectangular box with a Dirichlet boundary condition \cite{Canova:1992cj} and the electric field around a conducting sphere or cylinder \cite{Reitz}.
In fluid mechanics, we translate the problem to a potential flow around circular cylinder \cite{Batchelor}.

In this article, we pose the problem of a two-dimensional potential flow around a rectangle.
This is a modified version where a rectangle pole replaces a circular cylinder in standard problem set.
With this simple substitution the problem becomes analytically complex or intractable so that its exact solution is not known and that
it is rarely discussed in undergraduate curricula.
The problem can be solved using the conformal transformation \cite{Arfken, Jackson} whose prerequisite is complex analysis, an advanced subject in mathematics for undergraduates. 

We find an approximate, analytical solution of the problem by following three steps.
First, we find a solution in an infinite series of multipolar expansion.
Second, we apply the non-penetrating boundary condition by setting the vector component normal to the surface to be zero.
Third, we truncate the series and determine the coefficients of the series by using the multiple linear regression (MLR).

Introducing this approach to the physics undergraduate curricula, we expect the students to gain a conceptual understanding of superimposing multipoles.
The survey shows that students experience difficulties in understanding the physical concept of multipoles, primarily because they incorrectly regard the multipoles as substantial physical entities \cite{Grosse:1984gf, Kim:2016fq}.
By showing the detailed process of calculating a field under a specific Neumann boundary condition, our approach shows how the multipolar expansion is actually used and why the multipoles of higher order are introduced to match the given boundary conditions.
In addition, the presented method uses a multivariate modeling framework, which provides an excellent opportunity to employ simplification and think about major vs. minor components of a physical model, as well as to practice optimization for the estimation of coefficients derived via maximum likelihood method.

The educational practice is not limited to a specific application as long as the problem is properly modified according to the context. We set the problem in the context of fluid mechanics because of our familiarity: what is the potential function in a two-dimensional irrotational fluid flow around a rectangle? 
One can set up a similar problem in electrostatics because the mathematical formulations are exactly the same.

\section{Problem Statement}

A stream of incompressible and irrotational fluid flows in the positive $x$ direction at a mean speed $U$ and meets a rectangular pole of infinite length at the origin.
The flow is two-dimensional and does not penetrate the inner boundary of the pole, satisfying the slip boundary condition.
Find the scalar potential functions where (a) the flow faces one vertex and (b) the flow faces one side of the rectangle.

\section{Method}

\subsection*{Step 1: General Solution}

The potential function in the problem satisfies the Laplace equation, whose solution is very well known as an infinite series. 
The solution is a combination of radial and polar solutions, and  
in mathematical terms,
\begin{equation}
\phi  = U \left\{ \begin{aligned} r^{+m} \\ r^{-m} \end{aligned} \right\} \cdot \left\{ \begin{aligned} \cos{m\theta}\\ \sin{m\theta} \end{aligned} \right\},
\label{eq:general_sol}
\end{equation}
where we set $U=1$ without loss of generality.
We consider that i) the field far from the origin is to the $x$ direction, and ii) the potential satisfies two symmetry conditions one in $x$ and the other in $y$ direction. More precisely, the potential function is symmetric about the $x$-axis, $\phi(r,\theta)=\phi(r,-\theta)$, and antisymmetric about the $y$-axis, $\phi(r,\theta-\pi/2)=-\phi(r,-\theta+\pi/2)$.
These conditions reduce the general solution in Eq. (\ref{eq:general_sol}) to
\begin{equation}
\phi=r\cos\theta+\sum_{n=1}^{\infty}A_{n}r^{-(2n-1)}\cos\left[(2n-1)\theta\right].
\label{eq:general_sol_bc}
\end{equation}
The first term in Eq. (\ref{eq:general_sol_bc}) represents the mean flow far from the origin, and each term in the following summation, $r^{-(2n-1)}\cos\left[(2n-1)\theta\right]$, represents a multipole composed of $2n-1$ pairs of dipoles, which respectively have strength of $A_{n}$ arranged by the boundary condition.

\subsection*{Step 2: Boundary Equations}

First, consider the potential flow around a \textit{diamond} cylinder whose vertex meets the flow. 
In polar coordinates, the boundaries of the diamond are expressed as
\begin{equation}
r=\begin{cases}
{a}/{(\sin\theta+\cos\theta)} &\text{ for }0\leq \theta < \frac{\pi}{2}\\
{a}/{(\sin\theta-\cos\theta)} &\text{ for }\frac{\pi}{2}\leq \theta < \pi\\
{a}/{(-\sin\theta-\cos\theta)} &\text{ for }\pi \leq \theta < \frac{3 \pi}{2}\\
{a}/{(-\sin\theta+\cos\theta)} &\text{ for }\frac {3\pi}{2} \leq \theta < 2\pi,
\end{cases}
\label{eq:diamond_bound}
\end{equation}
where $a$ is one half of the diagonal. 
In the first quadrant, the Neumann boundary condition is given such that the field does not penetrate the boundary, i.e. $(\nabla\phi)\cdot\hat{n}=0$ on the boundary, where $\hat{n}=(\hat{x}+\hat{y})/\sqrt{2}$.
From Eq. (\ref{eq:general_sol_bc}), we find that
\begin{eqnarray}
v_n= \frac{1}{\sqrt{2}}-\sum_{n} \left( A_{n}\frac{2n-1}{r^{2n}}\cdot \frac{\cos2n\theta+\sin2n\theta}{\sqrt{2}}\right). 
\label{eq:diamond_vn}
\end{eqnarray}
Substituting $r=a/(\sin\theta+\cos\theta)$, Eq. (\ref{eq:diamond_vn}) becomes 
\begin{equation}
0 = -\frac{1}{\sqrt{2}}+\sum_{n}A_{n}\frac{2n-1}{ (a^2/2)^n} \sin^{2n}\left(\theta+\frac{\pi}{4}\right)\sin\left(2n\theta+\frac{\pi}{4}\right),
\label{eq:diamond_boundary_equation}
\end{equation}
where $0 \leq \theta<\pi/2$. Assume the side of a diamond is 2, which means $a = \sqrt{2}$. 
The step-by-step derivation of Eq. (\ref{eq:diamond_boundary_equation}) is in the Appendix.
In the second quadrant, we repeat the same calculation and get 
\begin{equation}
0 = \frac{1}{\sqrt{2}}+\sum_{n}A_{n}(2n-1)\sin^{2n}\left(\theta-\frac{\pi}{4}\right)\sin\left(2n\theta-\frac{\pi}{4}\right),
\label{eq:diamond_boundary_equation_q2}
\end{equation}
for $\pi/2 \leq \theta < \pi$.
Equations. (\ref{eq:diamond_boundary_equation}) and (\ref{eq:diamond_boundary_equation_q2}) are identical under the substitution $\theta=\pi-\theta'$.
This is expected because Eq. (\ref{eq:general_sol_bc}) assumes the antisymmetry about the $y$ axis.
Likewise, the boundary equations in the third and fourth quadrants are irrelevant, and therefore Eq. (\ref{eq:diamond_boundary_equation}) is the only equation to solve for to determine the coefficients $A_{n}$'s.


Next, consider the potential flow around a \textit{square} cylinder, where the flow faces one side of the rectangle.
In the first quadrant, the boundary of the square in polar coordinates is
\begin{equation}
r=\begin{cases}b\cos\theta &\text{ for }0\leq\theta<\frac{\pi}{4}\\
b\sin\theta &\text{ for }\frac{\pi}{4}\leq\theta<\frac{\pi}{2}, 
\end{cases}
\label{eq:square_bound}
\end{equation}
where $b$ is a half of the base length.
Applying the non-penetrating boundary condition, $\vec{v}\cdot\hat{n}=0$, we get
\begin{equation}
v_n=0=-1+\sum_{n}B_{n} \frac{2n-1}{b^{2n}} \cos^{2n}\theta\cos2n\theta
\label{eq:square_vn1}
\end{equation}
for $0\leq\theta<\pi/4$ and
\begin{equation}
v_n=0=\sum_{n}B_{n} \frac{2n-1}{b^{2n}} \sin^{2n}\theta\sin2n\theta
\label{eq:square_vn2}
\end{equation} 
for $\pi/4\leq\theta<\pi/2$.
Here, we use $B_n$ for the coefficients to distinguish the rectangle scenario from the diamond. 
For simplicity, we set the length of the side to 2, which means $b=1$. Merging Eqs. (\ref{eq:square_vn1}) and (\ref{eq:square_vn2}) using the Heaviside function $H$,
we get 
\begin{multline}
0 = -1+ H(\theta-\frac{\pi}{4})+\sum_{n} B_n\left(2n-1\right) \times \\
 \left\{ \left[1-H(\theta-\frac{\pi}{4})\right]\cos^{2n}\theta\cos2n\theta+H(\theta-\frac{\pi}{4})\sin^{2n}\theta\sin2n\theta\right\},
\label{eq:square_boundary_equation}
\end{multline}
where $0 \leq \theta<\pi/2$.

\subsection*{Step 3: Determine the Scale Using a Multiple Linear Regression}

We determine the scale of the coefficients $\{A_n\}$ and $\{B_n\}$'s of the boundary equations in (\ref{eq:diamond_boundary_equation}) and (\ref{eq:square_boundary_equation}) respectively using the multiple linear regression (MLR).
Because it is impracticable to obtain the solution with full precision, we find an optimal and approximate solution.
First, we truncate the infinite series and discretize the equation. We assume that a truncated finite series captures a large proportion of the physics and that the remaining higher order terms are considered to be part of systematic difference between theory and measurement.

Formally, we rewrite Eq. (\ref{eq:diamond_boundary_equation}) as
\begin{equation}
0 = -\frac{1}{\sqrt{2}}+ \sum_{n=1}^{N} A_{n} X_n (\theta) +\epsilon(\theta),
\label{eq:diamond_model}
\end{equation}
where $X_n(\theta)=(2n-1)\sin^{2n}\left(\theta+\frac{\pi}{4}\right)\sin\left(2n\theta+\frac{\pi}{4}\right)$ is the model driver, $N$ is the number of terms in the truncated series, and $\epsilon(\cdot)$ is the difference between the approximate solution and the exact solution. 

For a multiple regression, we prepare the driver (explainable factor) matrix $\mathbf{X}$:
\begin{equation}
\mathbf{X}=\left[ \mathbf{X}_1 \, \mathbf{X}_2 \,\cdots \, \mathbf{X}_N \right],
\end{equation}
where each column vector $\mathbf{X}_n$ is of length M
\begin{equation}
\mathbf{X}_n=\left[
\begin{array}{c}
X_n(\theta_1)\\
X_n(\theta_2)\\
\vdots\\
X_n(\theta_M)
\end{array}
\right],
\end{equation}
and it represents the driver value at fixed interval $\{ 0, \frac{\pi}{2M}, \cdots, \frac{(M-1)\pi}{2M} \}$. Therefore, $\mathbf{X}$ is an $M\times N$ matrix.
Rewriting Eq. (\ref{eq:diamond_model}) in a matrix form,
\begin{equation}
\mathbf{Y}=\mathbf{X}\mathbf{A}+\mathbf{\epsilon},
\label{eq:diamond_matrix_form}
\end{equation}
where $\mathbf{A}=[ A_1, A_2, \cdots, A_N]^T$, $\mathbf{\epsilon}=[\epsilon_1, \epsilon_2, \cdots, \epsilon_M]^T$ and $\mathbf{Y}$ is a constant vector of length $M$, 
\begin{equation}
\mathbf{Y}=\left[
\begin{array}{c}
1/\sqrt{2}\\
1/\sqrt{2}\\
\vdots\\
1/\sqrt{2}
\end{array}
\right].
\end{equation}

In multiple linear regression, we solve for $\mathbf{A}$ in Eq. (\ref{eq:diamond_matrix_form}) by minimizing the sum of squared differences \cite{Casella} between known $\mathbf{Y}$ and the approximate solution $\mathbf{X}\mathbf{A}$:
\begin{equation}
\arg\min_{\{A_n\}}\sum_m{\left(\frac{1}{\sqrt 2} - \sum_{n}X_n (\theta_m)A_n\right)^2}.
\end{equation}
Three important conditions for the multiple linear regression are i) the drivers of the regression are linear predictors of $\mathbf{Y}$, ii) $\mathbf{X}_n$'s are linearly independent from each other, and iii) constant variance in $\{\epsilon_n\}$ over the support. Then, the MLR solution provides a set of $\{A_n\}$ that are interpretable in physical sense. 
Formally, our coefficient estimates are obtained by
\begin{equation}
\mathbf{A}=(\mathbf{X}^T\mathbf{X})^{-1}\mathbf{X}^T\mathbf{Y}.
\end{equation}
Then the residual of the regression is
\begin{equation}
\mathbf\epsilon=\mathbf{Y}-\mathbf{X}(\mathbf{X}^T\mathbf{X})^{-1}\mathbf{X}^T\mathbf{Y}.
\end{equation}
For computation, we use the linear model function \textit{lm} in standard R distribution \cite{Team:2018aa}.

Again, we follow the same procedure to solve for the square case.
We discretize the continuous support in $M$ equal intervals and rewrite the boundary equation (\ref{eq:square_boundary_equation}) as
\begin{equation}
\mathbf{Y}^{(s)}= \mathbf{X}^{(s)}\mathbf{B}+\mathbf{\epsilon}^{(s)},
\label{eq:square_matrix_form}
\end{equation}
where the superscript {\it s} denotes a square case, $\mathbf{B}=[B_1, B_2, \cdots, B_N]^T$, $Y_m^{(s)}=1-H(\theta_m-\pi/4)$, or
\begin{equation}
\mathbf{Y}^{(s)}=\left[
\begin{array}{c}
1\\
\vdots\\
1\\
0\\
\vdots\\
0
\end{array}
\right]
\begin{array}{@{\kern-\nulldelimiterspace}l@{}}
	\begin{array}{@{}c@{}}\end{array}
	\begin{array}{@{}c@{}}\end{array}
  \end{array},
\end{equation}
and the $n^{th}$ column vector in the driver matrix $\mathbf{X^{{s}}}$ is
\begin{equation}
\mathbf{X}_n^{(s)}=\left[
\begin{array}{c}
(2n-1)\cos^{2n}\theta_1 \cos2n\theta_1 \\
\vdots\\
(2n-1)\cos^{2n}\theta_{M/2} \cos2n\theta_{M/2}\\
(2n-1)\sin^{2n}\theta_{M/2+1} \sin2n\theta_{M/2+1}\\
\vdots\\
(2n-1)\sin^{2n}\theta_M \sin2n\theta_M
\end{array}
\right].
\end{equation}
At the end, the model coefficients are obtained by 
\begin{equation}
\mathbf{B}=(\mathbf{X}^{(s)T}\mathbf{X}^{(s)})^{-1}\mathbf{X}^{(s)T}\mathbf{Y^{(s)}}.
\end{equation}

\section{Results}

We demonstrate the evaluation of the coefficients for the setting $M=1,000$ and $N=50$. 
In detail, the boundary equations are discretized into 1,000 evenly distributed points on one side of a rectangle.
We find that the result is not sensitive to the choice of $M$ as long as $M\gg N $.
We choose the multi-polar expansion to be represented using the the first fifty terms ($N=50$) of the infinite series because the difference between the theoretical boundary condition and the model estimate stabilizes around fitting the first fifty coefficients.

\begin{figure}
\begin{centering}
\includegraphics[width=8cm]{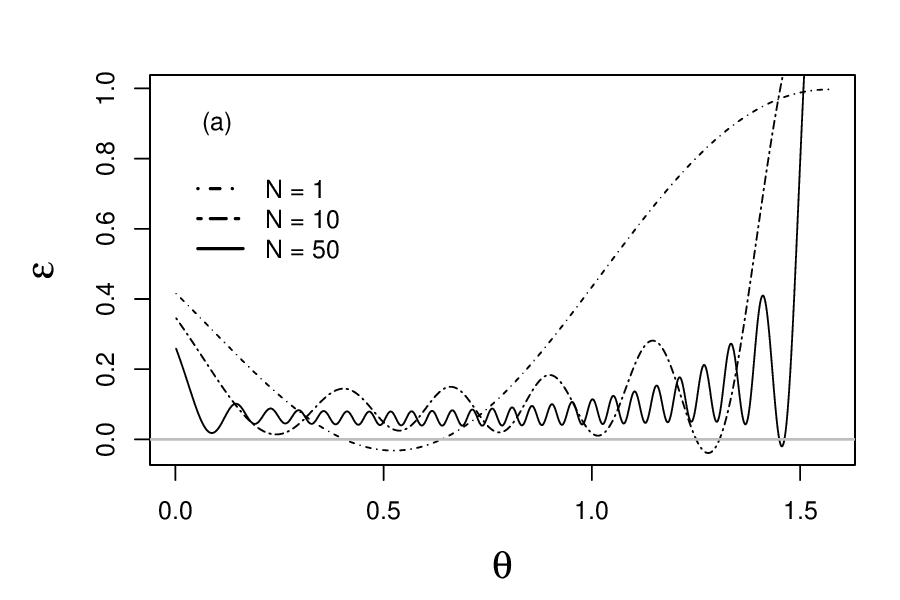}
\includegraphics[width=8cm]{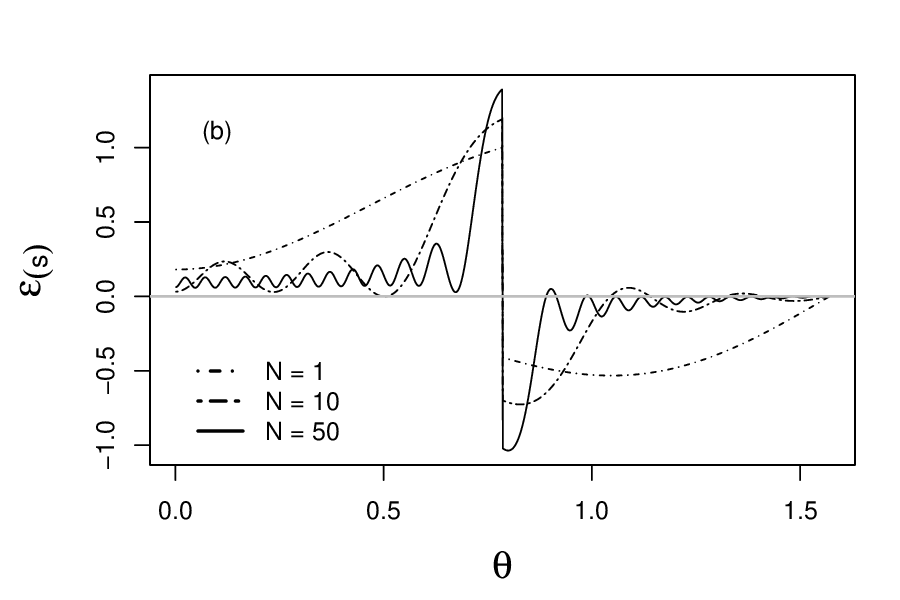}
\par
\end{centering}
\caption{ 
$\mathbf\epsilon$ vs. $\theta$ for (a) diamond and (b) square.
\label{fig:fitting}}
\end{figure}

In Fig. \ref{fig:fitting}(a), the fitted model difference $\mathbf{\epsilon}$ is plotted with respect to $\theta$ for the diamond case.
Three curves show the results of MLR for $N=1,\,10,\,50$.
As expected, the size of $\mathbf\epsilon$ gets smaller as the number of terms $N$ increases. 
When $N=50$, the discrepancy between the full and the fitted model flow is localized near the tips of the rectangle, where $\theta=0$ and $\theta=\pi/2$.
Mathematically, these are the singular points at which the surface normal is not well-defined, and therefore the discrepancy at the tips cannot be removed completely.
In Fig. \ref{fig:fitting}(b), the fitted model difference $\mathbf{\epsilon}^{(s)}$ is plotted for the square case. 
Similarly, the fit improves as $N$ increases, When $N=50$, we see that the discrepancy is localized at the tip of the rectangle, which is $\theta=\pi/4$.

Using the fitted values of the coefficients, the major analytic expression of the solution for a diamond is
\begin{equation}
\phi^{(d)}=r\cos\theta+1.19\frac{\cos\theta}{r}-0.23\frac{\cos3\theta}{r^3}+0.30\frac{\cos5\theta}{r^5} \cdots,
\label{eq:sol_diamond}
\end{equation}
and for square,
\begin{equation}
\phi^{(s)}=r\cos\theta+1.19\frac{\cos\theta}{r}+0.23\frac{\cos3\theta}{r^3}-0.30\frac{\cos5\theta}{r^5} \cdots
\label{eq:sol_square}
\end{equation}
The rest of coefficients are summarized in Tables \ref{tab:An} and \ref{tab:Bn}.
We note that $|A_n/B_n|\approx1$ for all $n$.
The coefficients $A_n$'s and $B_n$'s differ only by their sign, as both solutions are the rotational transformations of each other.

\begin{figure}
\begin{centering}
\includegraphics[viewport=75 65 410 395, clip=true, width=4cm]{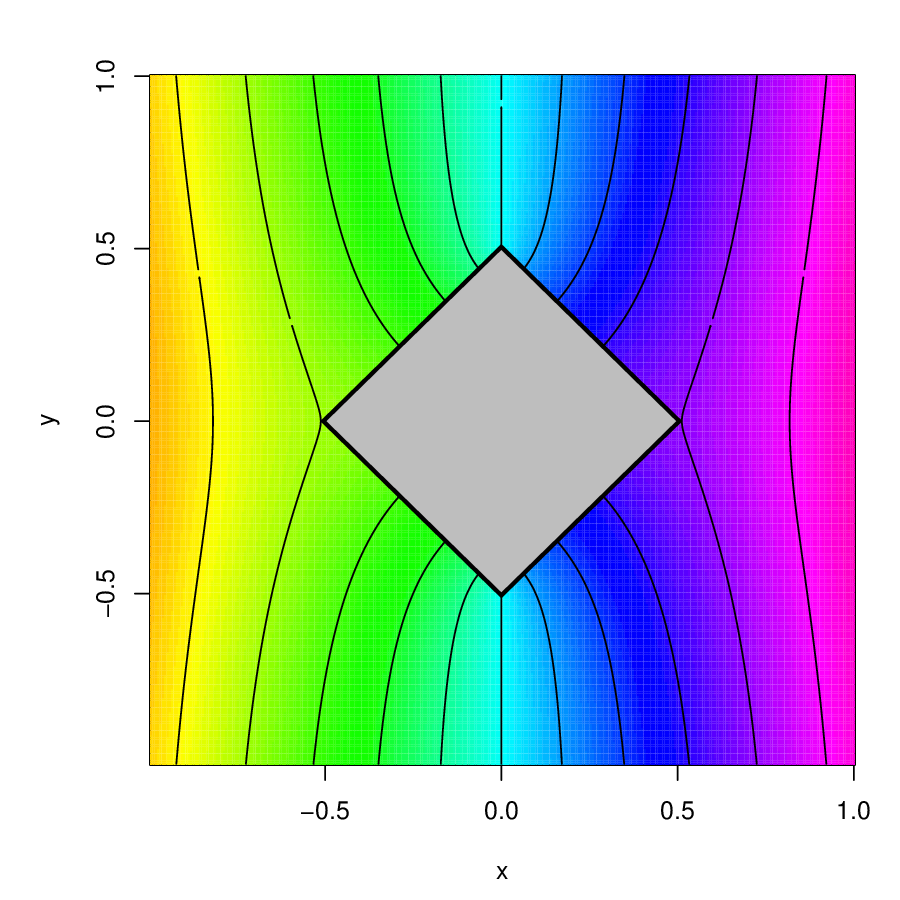}
\includegraphics[viewport=75 65 410 395, clip=true, width=4cm]{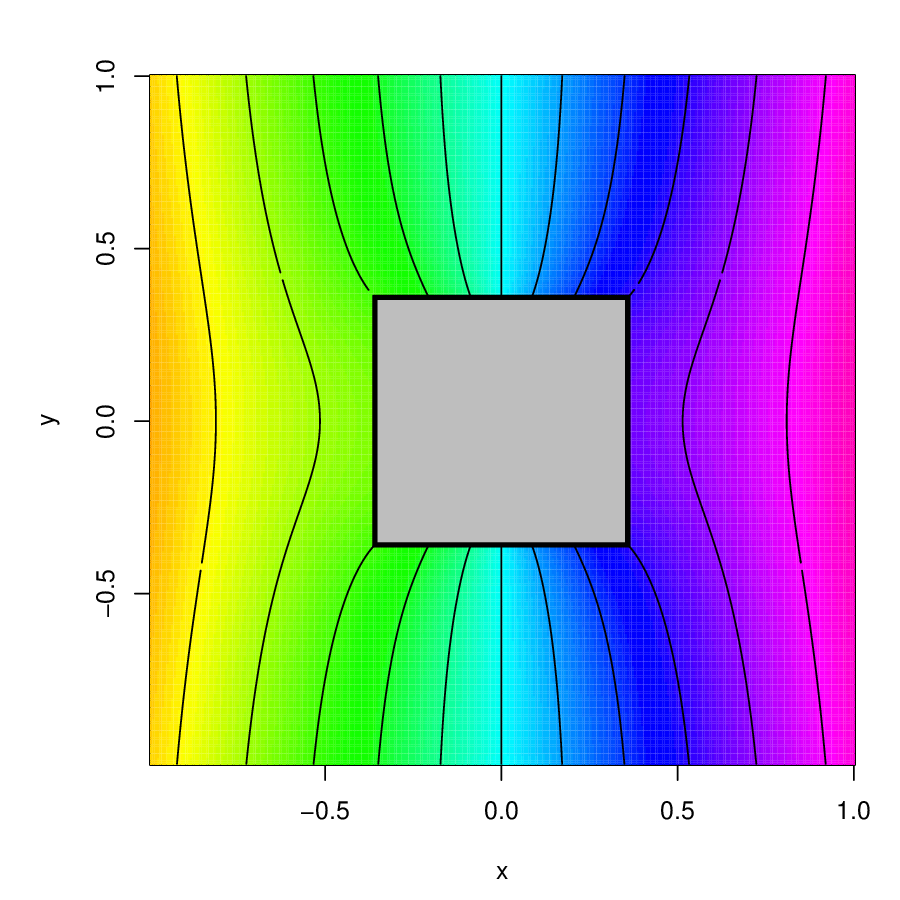}
\par
\end{centering}
\caption{ 
Equipotential lines for (a) diamond and (b) square. 
The length of the side of rectangles is 2.
\label{fig:equipotential_lines}}
\end{figure}

Using the solutions acquired from MLR, we plot the equipotential contours in Fig. \ref{fig:equipotential_lines}.
It is shown that the equipotential lines are approximately perpendicular to the boundary at the center,
and therefore the vector field representing the gradient of the potential function has no normal component at the boundaries.

\begin{table}[h]
\centering
\begin{tabular}{c|c||c|c||c|c||c|c||c|c}
\hline
$A_{1}$ & 1.19 & $A_{2}$ & -0.23 & $A_{3}$ & 0.30 & $A_{4}$ & -0.20 & $A_{5}$ & 0.35 \\
$A_{6}$ & -0.31 & $A_{7}$ & 0.57 & $A_{8}$ & -0.56 & $A_{9}$ & 1.05 & $A_{10}$ & -1.10 \\
$A_{11}$ & 2.03 & $A_{12}$ & -2.20 & $A_{13}$ & 4.00 & $A_{14}$ & -4.38 & $A_{15}$ & 7.80 \\
$A_{16}$ & -8.59 & $A_{17}$ & 14.8 & $A_{18}$ & -16.3 & $A_{19}$ & 27.2 & $A_{20}$ & -29.8 \\
$A_{21}$ & 47.6 & $A_{22}$ & -51.7 & $A_{23}$ & 78.8 & $A_{24}$ & -84.7 & $A_{25}$ & 122 \\
$A_{26}$ & -130 & $A_{27}$ & 177 & $A_{28}$ & -186 & $A_{29}$ & 236 & $A_{30}$ & -244 \\
$A_{31}$ & 287 & $A_{32}$ & -292 & $A_{33}$ & 315 & $A_{34}$ & -316 & $A_{35}$ & 308 \\
$A_{36}$ & -305 & $A_{37}$ & 264 & $A_{38}$ & -257 & $A_{39}$ & 194 & $A_{40}$ & -186 \\
$A_{41}$ & 119 & $A_{42}$ & -112 & $A_{43}$ & 59.1 & $A_{44}$ & -53.7 & $A_{45}$ & 21.3 \\
$A_{46}$ & -19.3 & $A_{47}$ & 5.19 & $A_{48}$ & -4.63 & $A_{49}$ & 0.630 & $A_{50}$ & -0.553 \\
\hline
\end{tabular}
\caption{The values of $A_n$'s for the 2-D potential flow around a diamond}
\label{tab:An}
\end{table}

\begin{table}[h]
\centering
\begin{tabular}{c|c||c|c||c|c||c|c||c|c}
\hline
$B_{1}$ & 1.19 & $B_{2}$ & 0.23 & $B_{3}$ & -0.30 & $B_{4}$ & -0.20 & $B_{5}$ & 0.35 \\
$B_{6}$ & 0.31 & $B_{7}$ & -0.57 & $B_{8}$ & -0.56 & $B_{9}$ & 1.04 & $B_{10}$ & 1.09 \\
$B_{11}$ & -2.02 & $B_{12}$ & -2.17 & $B_{13}$ & 3.97 & $B_{14}$ & 4.34 & $B_{15}$ & -7.75 \\
$B_{16}$ & -8.50 & $B_{17}$ & 14.7 & $B_{18}$ & 16.1 & $B_{19}$ & -27.0 & $B_{20}$ & -29.4 \\
$B_{21}$ & 47.2 & $B_{22}$ & 51.1 & $B_{23}$ & -78.2 & $B_{24}$ & -83.7 & $B_{25}$ & 122 \\
$B_{26}$ & 129 & $B_{27}$ & -176 & $B_{28}$ & -183 & $B_{29}$ & 234 & $B_{30}$ & 241 \\
$B_{31}$ & -285 & $B_{32}$ & -289 & $B_{33}$ & 313 & $B_{34}$ & 312 & $B_{35}$ & -306 \\
$B_{36}$ & -301 & $B_{37}$ & 262 & $B_{38}$ & 254 & $B_{39}$ & -193 & $B_{40}$ & -183 \\
$B_{41}$ & 118 & $B_{42}$ & 110 & $B_{43}$ & -57.6 & $B_{44}$ & -53.0 & $B_{45}$ & 21.1 \\
$B_{46}$ & 19.1 & $B_{47}$ & -5.14 & $B_{48}$ & -4.56 & $B_{49}$ & 0.624 & $B_{50}$ & 0.545 \\
\hline
\end{tabular}
\caption{The values of $B_n$'s for the 2-D potential flow around a square}
\label{tab:Bn}
\end{table}

\section{Discussion}
\label{sec:discussion}


Conceptually, the proposed procedure can be interpreted as an analysis of data acquired from a thought experiment.
In this thought experiment, suppose we have a probe that measures the flow speed at the boundary of the rectangle.
Assuming the flow is incompressible and irrotational, the measurement should be zero at the boundary.
The full solution in Eq. (\ref{eq:sol_diamond}) is an outcome of regression analysis based on the mathematical framework in Eq. (\ref{eq:diamond_model}) and the thought experiment data.

We deem this material to be worthy of discussion in upper-level undergraduate classes.
The primary educational goal is to understand the multipolar expansion and to dispel misunderstanding of the physical substantiality of the multipoles.
When the potential flow around a circle (or the electric field around a metallic cylinder in electrostatics) is presented, the solution is exact and contains only the dipole flow. When students are exposed to higher order multipoles that never appear in the soluble case, we bring a rich understanding of the Laplace equation and present a flexible visual imagery of the potential flow. 
The second educational goal is to practice the regression analysis.
While the least square fitting is frequently applied in the experimental courses of physics, the direct translation of physical phenomena to conceptual, analytical platform is rarely done in the physics curricula. The simple principle/idea of statistics is to present a phenomenon in two parts -- the explainable and the unexplainable, such as measurement errors or small, ignorable differences.
The current method provides an opportunity for such discussion. The hands-on experience of finding an optimal strength of the higher order multipoles, students are expected to learn the poles are virtual sources and sinks that is engineered to match a specific boundary condition.

In fluid mechanics, the singular perturbative nature of the governing equation \cite{Neu2015} leads the separation of the scales.
Therefore, the analytic expression of the potential flow may provide an insight for some problems \cite{Kim:2015jp}, where the flow far from the boundary is concerned.


\appendix 

\section{Step-by-step derivation of Eq. (\ref{eq:diamond_boundary_equation})}

The velocity field is calculated from the general solution of the potential function in Eq. (\ref{eq:general_sol_bc}). 
\begin{eqnarray}
\vec{v}&=&\hat{r} \frac{\partial\phi}{\partial{r}}+\hat{\theta}\frac{1}{r} \frac{\partial\phi}{\partial\theta},
\end{eqnarray}
where the radial component of the velocity field 
\begin{equation}
v_r= \cos\theta- \sum_n A_n r^{-2n} \cos[(2n-1)\theta] \cdot (2n-1)
\end{equation}
and the azimuthal component 
\begin{equation}
v_\theta=-\sin\theta - \sum_n A_n r^{-2n} \sin[(2n-1)\theta] \cdot (2n-1)
\end{equation} 
Using the surface normal $\hat{n}=({\hat{x}+\hat{y}})/{\sqrt{2}}$, the normal velocity is
\begin{equation}
v_n= \vec{v}\cdot\hat{n}=v_r\left(\frac{\hat{r}\cdot\hat{x}+\hat{r}\cdot\hat{y}}{\sqrt{2}}\right)+v_\theta\left(\frac{\hat\theta\cdot\hat{x}+\hat\theta\cdot\hat{y}}{\sqrt{2}}\right).
\end{equation}
Using $\hat{r}\cdot\hat{x}=\cos\theta$, $\hat{r}\cdot\hat{y}=\sin\theta$, $\hat{\theta}\cdot\hat{x}=-\sin\theta$, and $\hat{\theta}\cdot\hat{y}=\cos\theta$,
\begin{eqnarray}
\sqrt{2}v_n &=&  \cos\theta(\cos\theta+\sin\theta)-\sin\theta(-\sin\theta+\cos\theta) \nonumber\\&& + \sum_n A_n (2n-1) r^{-2n} P \\
&=&1+\sum_n A_n (2n-1) r^{-2n} P,
\end{eqnarray}
where
\begin{eqnarray}
P&=&-\cos[(2n-1)\theta]\cos\theta-\cos[(2n-1)\theta]\sin\theta \nonumber \\
&&+\sin[(2n-1)\theta\sin\theta-\sin[(2n-1)\theta]\cos\theta.
\end{eqnarray}
Using the trigonometric identities $\sin\alpha\cos\beta+\cos\alpha\sin\beta=\sin(\alpha+\beta)$ and $\cos\alpha\cos\beta-\sin\alpha\sin\beta=\cos(\alpha+\beta)$, it follows 
\begin{eqnarray}
P&=&-\cos[(2n)\theta]- \sin[(2n)\theta].
\end{eqnarray}
From the above results, Eq. (\ref{eq:diamond_vn}) is derived,
\begin{eqnarray}
v_n = \frac{1}{\sqrt{2}} - \sum_n A_n \frac{2n-1}{r^{2n}} \frac{\cos2n\theta+\sin2n\theta}{\sqrt{2}}.
\end{eqnarray}
We now substitute $r=a/(\cos\theta+\sin\theta)$, 
\begin{eqnarray}
v_n = \frac{1}{\sqrt{2}} - \sum_n A_n\frac{2n-1}{a^{2n}} \frac{Q}{\sqrt{2}},
\end{eqnarray}
where
\begin{eqnarray}
Q=\left( \cos\theta+\sin\theta\right)^{2n} \left( \cos2n\theta+\sin2n\theta\right).
\end{eqnarray}
Using $\sin\theta\cos\pi/4+\sin\pi/4\cos\theta=\sqrt{2}^{-1}(\sin\theta+\cos\theta)=\sin(\theta+\pi/4)$,
\begin{eqnarray}
Q=  \sqrt{2}^{2n+1} \sin^{2n}\left(\theta+\frac{\pi}{4}\right) \sin \left(2n\theta+\frac{\pi}{4} \right) .
\end{eqnarray}
Putting $Q$ back to $v_n$,
\begin{eqnarray}
v_n = \frac{1}{\sqrt{2}} - \sum_n A_n\frac{2n-1}{(a^2/2)^n}  \sin^{2n}\left(\theta+\frac{\pi}{4}\right) \sin \left(2n\theta+\frac{\pi}{4} \right).
\end{eqnarray}
We further simplify the equation by setting $a^2/2=1$.
This specification concludes the derivation of Eq. (\ref{eq:diamond_boundary_equation}).


\bibliography{potentialflow2014}

\end{document}